\documentclass[preprint2]{aastex}

\shorttitle{High Contrast Imaging with a PIAA Coronagraph}
\shortauthors{Guyon et al.}
\usepackage{graphicx}
\usepackage[mathscr]{eucal}
\begin{document}

\title{High Contrast Imaging and Wavefront Control with a PIAA Coronagraph: Laboratory System Validation}

\author{Olivier Guyon}
\affil{National Astronomical Observatory of Japan, Subaru Telescope, Hilo, HI 96720}
\affil{Steward Observatory, University of Arizona, Tucson, AZ 85721}
\email{guyon@naoj.org}

\author{Eugene Pluzhnik}
\affil{NASA Ames Research Center, Mountain View, CA 94035}

\author{Frantz Martinache, Julien Totems, Shinichiro Tanaka}
\affil{National Astronomical Observatory of Japan, Subaru Telescope, Hilo, HI 96720}

\author{Taro Matsuo}
\affil{NASA Jet Propulsion Laboratory, Pasadena, CA 91109}

\author{Celia Blain}
\affil{University of Victoria, Victoria, BC, Canada V8W 3P6}

\author{Ruslan Belikov}
\affil{NASA Ames Research Center}

\begin{abstract}

The Phase-Induced Amplitude Apodization (PIAA) coronagraph is a high performance coronagraph concept able to work at small angular separation with little loss in throughput. We present results obtained with a laboratory PIAA system including active wavefront control. The system has a 94.3\% throughput (excluding coating losses) and operates in air with monochromatic light.

Our testbed achieved a 2.27 $10^{-7}$ raw contrast between 1.65 $\lambda/D$ (inner working angle of the coronagraph configuration tested) and 4.4 $\lambda/D$ (outer working angle). Through careful calibration, we were able to separate this residual light into a dynamic coherent component (turbulence, vibrations) at 4.5 $10^{-8}$ contrast and a static incoherent component (ghosts and/or polarization missmatch) at 1.6 $10^{-7}$ contrast. Pointing errors are controlled at the $10^{-3}$ $\lambda/D$ level using a dedicated low order wavefront sensor. 

While not sufficient for direct imaging of Earth-like planets from space, the  2.27 $10^{-7}$ raw contrast achieved already exceeds requirements for a ground-based Extreme Adaptive Optics system aimed at direct detection of more massive exoplanets. We show that over a 4hr long period, averaged wavefront errors have been controlled to the 3.5 $10^{-9}$ contrast level. This result is particularly encouraging for ground based Extreme-AO systems relying on long term stability and absence of static wavefront errors to recover planets much fainter than the fast boiling speckle halo.
\end{abstract}
\keywords{instrumentation: adaptive optics --- techniques: high angular resolution}

\section{Introduction}
\label{sec:intro}
An imaging system aimed at dection or characterization (spectroscopy) of exoplanets must overcome the large contrast beween the planet and its star. This is particularly challenging for Earth-like planets, where the contrast is $\approx 10^{-10}$ in the visible and the angular separation is 0.1\arcsec for a system at 10pc. Many coronagraph concepts have recently been proposed to overcome this challenge (see review by \cite{guyo06}). Among the approaches suggested, Phase-Induced Amplitude Apodization (PIAA) coronagraphy is particularly attractive. In a PIAA coronagraph, aspheric optics (mirrors or lenses) apodize the telescope beam with no loss in throughput. A PIAA coronagraph combines high throughput, small inner working angle (2 $\lambda/D$ for $10^{-10}$ contrast), low chromaticity (when mirrors are used), full 360 degree discovery space, and full $1 \lambda/D$ angular resolution. Angular resolution (size of the planet's PSF in the image) is a critical performance parameter as exoplanet imaging sensitivity, even if speckles have been perfectly removed, is usually background-limited due to sky or thermal emission (near-IR and mid-IR imaging from the ground) or zodiacal and exozodiacal light (direct imaging of Earth-like planets in the visible from space).  
The PIAA concept, orginally formulated by \cite{guyo03}, has since been studied in depth in several subsequent publications \citep{trau03,guyo05,vand05,gali05,mart06,vand06,pluz06,guyo06,beli06,guyo09,lozi09}, which the reader can refer to for detailed technical information.

In the first laboratory demonstration of the PIAA concept \citep{gali05}, lossless beam apodization was demonstrated, and the field aberrations introduced by the PIAA optics were confirmed experimentally. In this first prototype, the PIAA acrilic optics lacked surface accuracy required for high contrast imaging, and since this experiment did not include active wavefront control, the high contrast imaging potential of the technique could not be demonstrated. In the present paper, we report on results obtained with a new system which includes reflective PIAA optics and wavefront control. Our prototype combines the main elements/subsystems envisionned for a successful PIAA imaging coronagraph instrument, with the exception of corrective optics required to remove the strong off-axis aberrations introduced by the PIAA optics. This last subsystem has been designed and built for another testbed, and its laboratory performance is reported in a separate paper \citep{lozi09}.

The overall system architecture adopted for our test is presented and justified in \S\ref{sec:archi}. The design of the main components of the coronagraphs (PIAA mirrors, masks) is also described in this section. Wavefront control and calibration are discussed in \S\ref{sec:wfc}. Laboratory results are presented in \S\ref{sec:labresults}.

\begin{figure*}[htb]
\includegraphics[scale=0.6]{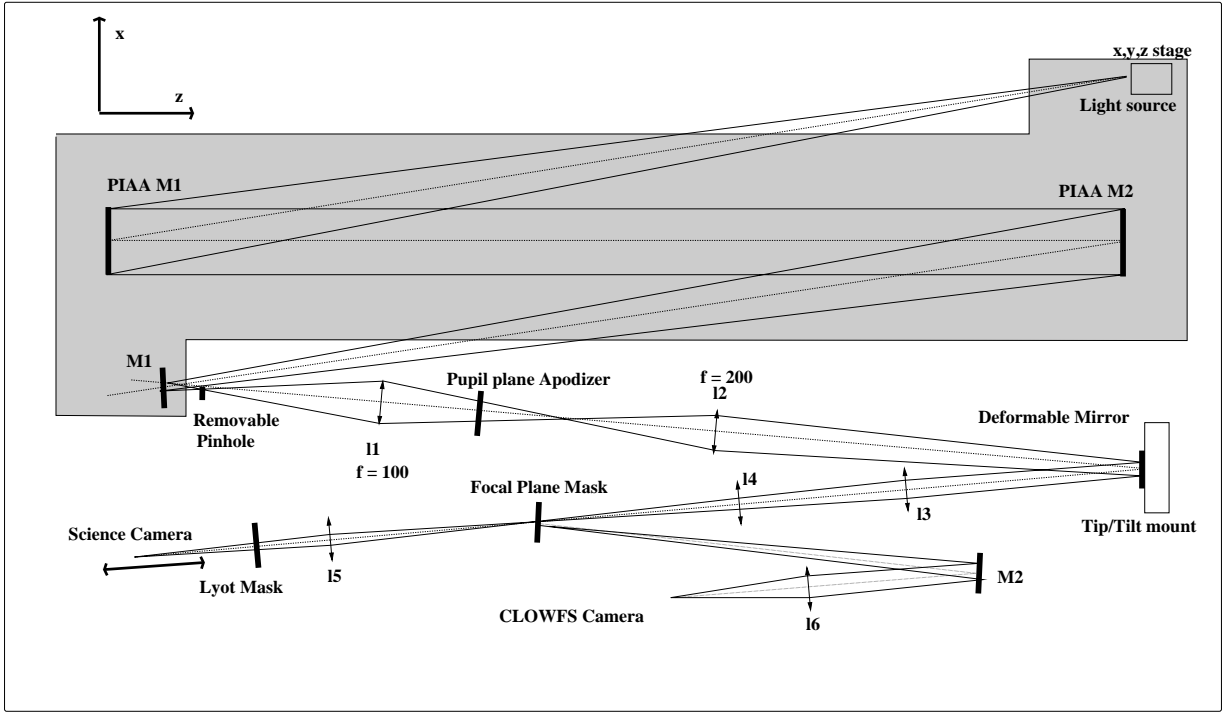} 
\caption{\label{fig:optlayout} Optical layout of the laboratory PIAA coronagraph system. The grey shaded area shows the rigid PIAA bench on which the two PIAA mirrors are mounted. The light source is at the upper corner of the figure. The focal plane mask (near the bottom, center) separates light into the imaging channel and the coronagraphic low order wavefront sensor (CLOWFS) channel.}
\end{figure*}

\section{Laboratory system architecture}
\label{sec:archi}

\subsection{Plate scale in a PIAA system}
Two optical conventions exist to define plate scale (physical distance (in meter) on the focal plane per unit of angle on the sky (in radian)): one is to follow the trajectory of the central ray in the optical system and measure its displacement on the focal plane as the source is moved on the sky, and the other one is to define the plate scale as equal to $(F/D_{fp}) \times D$, where D is the telescope diameter and $F/D_{fp}$ is the beam F-ratio at the focal plane. While the two conventions give identical results in conventional optical systems, they yield different values at the focal plane after the PIAA optics. This effect, due to the pupil distortion introduced by the PIAA optics, is well documented in previous PIAA-related publications, and leads to some confusion when comparing plate scale values as both definitions have previously been used. In this paper, we chose to avoid using either definition; instead, we adopt a convention where focal plane scale is physically defined relative to sky angle: 1 $\lambda$/D in the focal plane is defined by the physical distance by which the PSF photocenter moves when the source position on the sky is offset by 1 $\lambda$/D.

\subsection{Coronagraph architecture}

The coronagraph architecture adopted is a hybrid PIAA \citep{pluz06}, where beam apodization is shared between the aspheric PIAA mirrors (described in \S\ref{ssec:PIAAmirrors}) and a post-apodizer (described in \S\ref{ssec:apodizer}). The PIAA mirrors perform most of the apodization, but leave a small amount of excess light at the edge of beam (left at 0.85\% of the surface brightness at the center of the beam), which is then removed by the apodizer. Thanks to this hybrid approach, the PIAA mirrors are more easily manufacturable (less aspheric) and the apodizer tolerances are relaxed (the apodizer is not absorbing light in the bright parts of the beam). The hybrid design also solves the problem of propagation-induced chromaticity \citep{vand06}, which would otherwise limit contrast at $\approx 10^{-7}$ in a non-hybrid system working in a 20\% wide band. While this second benefit was not relevant in our monochromatic experiment, it is key to enable high contrast direct imaging of exoplanet from space. 

The cost in throughput and angular resolution due to the apodizer are small since the apodizer only removes light in the fainter edges of the remapped beam.

A high contrast image is formed after the apodizer, where starlight is blocked by the focal plane mask. Since the upstream PIAA optics + apodizer have apodized the beam with little loss in telescope angular resolution, the focal plane mask is small, with a radius ranging from approximately 1 $\lambda/D$ on the sky for a $10^{-6}$ contrast goal to approximately 2 $\lambda/D$ on the sky for a $10^{-10}$ contrast goal. The focal plane mask is also part of the low order wavefront sensor (Guyon et al. 2009) briefly described in \S\ref{ssec:lowfs} which uses starlight reflected by the focal plane mask for accurate sensing of pointing errors and defocus.

The optical layout of the laboratory experiment is shown in Figure \ref{fig:optlayout}. The light source is a single mode fiber fed by a HeNe laser ($\lambda$ = 632.58 nm), mounted on a x,y,z stage for control of the input tip/tilt and focus. The PIAA system (mirrors PIAA M1 and PIAA M2) creates a converging apodized beam. PIAA M2 is chosen as the pupil plane for the system, and lens l1 creates a small image of the pupil plane onto the apodizer. Lens l2 reimages the pupil plane on the deformable mirror, which is located in a $\approx$F/60 converging beam. Lenses l3 and l4 form a focal plane for the focal plane mask. The reimaging lens l5 is used to create a pupil plane and a focal plane. A Lyot mask is located in the pupil plane, but can be remotely moved out to allow the science camera (which is nominally in the focal plane) to move forward and acquire a direct pupil plane image. The focal plane mask reflects some of the light to the coronagraphic low order wavefront sensor (CLOWFS) camera. A detailed description of this device is given in \cite{guyo09}.

Once the number of lenses/mirrors and their relative positions was decided, a set of equations was coded to link together the exact location of optical elements, their focal lengths (for lenses), the position of pupil and focal planes, the beam size on each optical element, and the plate scale in the focal planes. The equations contain all the hard constraints of the experiment (for example, the DM must be conjugated to the pupil plane apodizer). Each variable was given an allowed range (for example beam size on the DM) or a set of allowed values (focal lengths of lenses constrained by what is available from vendors). A randm search algorithm was then used to test many possible optical designs and select the solutions which meet the criteria. This approach provided us with a flexible tool to explore design options. The same optimization code was also used to compute offsets in the position of several components during fine alignment of the system.

\subsection{Wavefront control hardware and architecture}
\label{wfc:hardarch}
A single 32 by 32 actuators MEMS deformable mirror (DM) is used for wavefront control and is located after the PIAA optics. The wavefront control subsystem is therefore fully decoupled from beam shaping effects introduced by the PIAA optics. The PIAA optics simply deliver an apodized beam to the wavefront control subsystem, which operates independently of the technique used to apodize the beam (remapping vs. conventional apodization). While this configuration is simpler than a configuration where the DM is ahead of the PIAA optics for wavefront control, we note that it does not offer as wide a field of view due to the magnification effect described in \cite{guyo05}.

Our laboratory demonstration was performed in monochromatic light. In this configuration, a single DM provides sufficient degrees of freedom to remove coherent diffracted light in one half of the field of view, regardless of how phase and amplitude aberrations in the beam are created. In a real coronagraphic instrument operating in broadband light, diffractive propagation between optics needs to be taken into account when designing the wavefront control architecture: manufacturing errors on optics introduce amplitude errors and wavefront chromaticity. In a high contrast instrument, such errors can only be addressed with a multiple-DM configuration, where the DM locations are optimized to reduce residual wavefront errors over the spectral band used. Diffraction propagation effects between the aspheric PIAA optics surface would therefore need to be quantified when designing the wavefront control hardware.

Two challenging aspects of wavefront control in PIAA coronagraphs (understanding how beam shaping affects wavefront control, and polychromatic wavefront control) are therefore not addressed in our experiment.

\subsection{Aspheric PIAA mirror design and fabrication}
\label{ssec:PIAAmirrors}

\begin{figure*}[htb]
\includegraphics[scale=0.32]{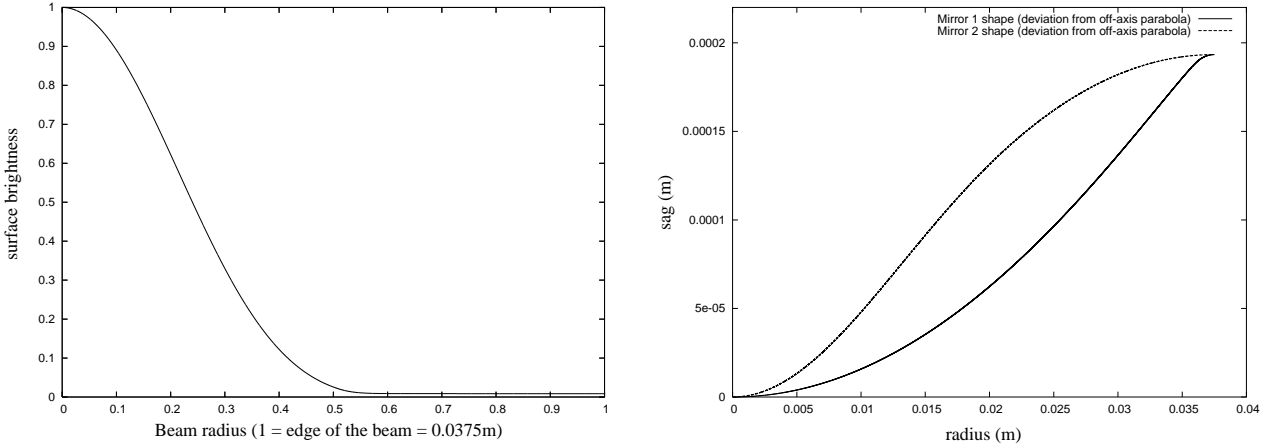}
\caption{\label{fig:M1M2shape} Left: Beam apodization profile used for the design of the PIAA system. In this hybrid design, the apodization is not complete, and the beam surface brightness at the edge of the beam is left at 0.85\% of the center surface brightness. This extra light will need to be removed by a conventional apodizer. Right: Apodization radial sag term for PIAA M1 and PIAA M2 in an on-axis configuration. The narrow region at the edge of PIAA M1 has a strong localized curvature, and is the most challenging feature for manufacturing the PIAA system optics.}
\end{figure*}

The geometric remapping is performed by two highly aspheric mirrors, the PIAA mirrors. The role of the first mirror is mostly to project on the second mirror the desired amplitude profile, which is partially apodized with a faint plateau on the outside of the beam (see Figure \ref{fig:M1M2shape}, left). The apodization profile is described in more detail in \cite{pluz06}, which includes a chromatic diffraction analysis of the PIAA optics used for our experiment. The PIAA M1 mirror acts as a converging element in the center (to concentrate more light in the center of the beam on PIAA M2) while light in the outside is diluted in a wide area of the beam on PIAA M2. This behavior explains the peculiar aspheric sag shown in Figure \ref{fig:M1M2shape} on the right. The PIAA M2 mirror's role is to re-collimate light to output a beam which is apodized but free of phase aberrations.

PIAA mirrors can be designed by solving a relatively simple differential equation when the input and output beams are collimated (the equation is given in \cite{guyo03}, and also in \cite{trau03} in a different form) or when the system is on-axis. In our laboratory experiment, the PIAA mirrors are focusing elements and the aspheric remapping shapes are added to off-axis parabolas. In this configuration, the PIAA mirror shapes cannot be derived from a simple differential equation, and they were designed by an iterative algorithm:
\begin{enumerate}
\item{Initialization: The PIAA mirror shapes are computed by solving the differential equation for an on-axis system.}
\item{A constant slope is added to each of the PIAA mirror. If there were no apodization, the mirrors obtained in step 1 would be on-axis parabolas, and this slope would turn them into off-axis parabolas. In the PIAA system, however, adding this slope only leads to an approximation to the off-axis PIAA system}
\item{A 3 dimensional raytracing code is used to compute the beam phase and amplitude on the surface of a sphere centered on the output focus of the system immediately after reflection on PIAA M2.}
\item{The difference between the measured and desired beam amplitude on the sphere is used to update PIAA M1's shape by linear decomposition of this residual (using pre-computed residuals obtained by adding Zernike polynomials on PIAA M1's shape).}
\item{The residual phase error measured on the sphere is compensated by changing PIAA M2's shape.}
\item{Return to Step 3 with the new mirror shapes.}
\end{enumerate}
This algorithm converges because changing PIAA M2's shape has little effect on the amplitude profile of the beam on the sphere, which is almost entirely a function of PIAA M1's shape.  

The PIAA shapes were computed for a 75mm beam diameter at the PIAA mirrors, a 1.125m separation from the center of PIAA M1 to the center of PIAA M2, and a 190mm offset between the PIAA M1 to PIAA M2 centerline and the input and output of the PIAA system (see Figure \ref{fig:optlayout}). In the coordinate system shown in Figure \ref{fig:optlayout}, each mirror shape can be written as:
\begin{equation}
z(x,y) = OAP(x,y) + sag(r) + \Sigma_i^j \alpha_i Z_i^j
\end{equation}

$OAP(x,y)$ is the off-axis parabola which would be the PIAA mirror shapes if no apodization was performed. It is a 1133 mm focal length OAP with a 190 mm off-axis distance from the center of the optical element. This shape is identical for the two PIAA mirrors although the orientation is different. $sag(r)$ is the apodization radial sag on each mirror. It is computed for an on-axis system. A corrective term is added to account for the fact that the system is off-axis (tends to 0 for an on-axis system),  decomposed as a sum of Zernikes polynomials up to radial order 7. For both PIAA mirrors, this correction is $\approx$95 nm RMS (excluding tip-tilt).

The most challenging feature of the system is the small radius of curvature in the outer part of the beam on PIAA M1. While the hybrid design adopted mitigates this problem, the radius of curvature still reaches a minimum of 155 mm near the edge of the mirror, at 36.1mm from the center of the mirror.
 
The PIAA mirrors were fabricated by Axsys Imaging Technologies. The mirror substrates were initially diamond turned according to the 3-D prescriptions described above, and then polished against computer generated  holograms (CGHs). PIAA M1 and PIAA M2 were then assembled on a rigid aluminum bench, aligned and permanently fixed to the bench. The residual system wavefront error was then reduced to 0.04 waves RMS by figuring PIAA M2. Two sets of PIAA mirrors (4 mirrors total) were manufactured.

\subsection{Post-apodizer and system throughput}
\label{ssec:apodizer}
\begin{figure*}[htb]
\includegraphics[scale=0.42]{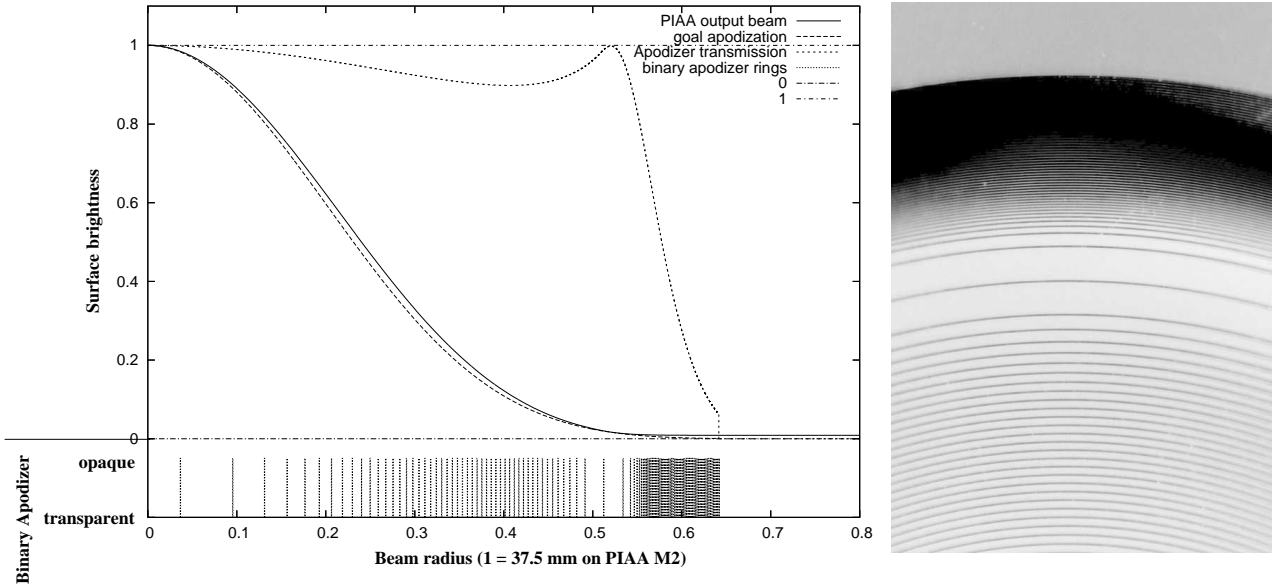} 
\caption{\label{fig:apodesign} Left: Apodizer design. Right: Microscope image of the outer part of the apodizer (the image covers approximately 1mm vertically). The outer edge of the apodizer (last transmissive ring) is at 1.9mm radius. The width of the individual opaque rings is 0.8$\mu m$.}
\end{figure*}

In order to ease manufacturing, the PIAA optics were designed to perform most, but not all, of the beam apodization required for high contrast imaging. A more conventional apodizing scheme is therefore necessary to transform the beam profile at the output of the PIAA optics (solid curve in Figure \ref{fig:apodesign}) into the desired beam profile (dashed curve in Figure \ref{fig:apodesign}).

The post apodizer was designed in transmission, with a series of narrow opaque rings blocking light. The position and width of the rings is optimized to best approximate the ideal continuous apodization profile shown in Figure \ref{fig:apodesign} as the curve labeled ``Apodizer transmission''. Several contraints were imposed on the design to ensure manufacturability: no ring should be less than 0.8 $\mu m$ wide and the gap between consecutive opaque rings should be no less than 5 $\mu m$. The resulting design is composed of 109 opaque rings for a total apodizer diameter of 3.815 mm (defined by the outer edge of the last opening between opaque rings). The apodizer was manufactured by lithography on a transmissive substrate.

The post-apodizer throughput over the 3.815mm diameter is 96.9 \%, but due to the narrow rings in the apodizer, some of the light transmitted is diffracted at large angles. The effective throughput of the apodizer is 94.3\%, and would be equal to the throughput if the apodizer were continuous instead of binary. The full system throughput can therefore reach 94.3\% (excluding losses due to coating) provided that the telescope pupil size on PIAA M1 is adjusted to the apodizer diameter. In practice, the telescope pupil should however be made slightly larger to allow for pupil centering errors, and in very high contrast applications (space coronagraphy), to mitigate possible edge ringing effects due to Fresnel propagation.

\begin{figure}[htb]
\includegraphics[scale=0.33]{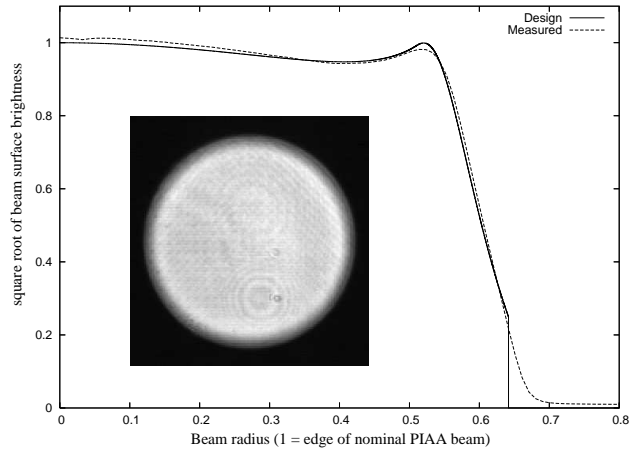} 
\caption{\label{fig:apodproflab} Laboratory measurement of the apodizer throughput. The designed and measured radial throughputs are compared.}
\end{figure}

The apodizer throughput was mesured by inserting a pinhole in the PIAA output focus and moving the science camera in the pupil plane. The 1 $\mu$m pinhole is used to de-apodize the beam at the expense of a very low throughput. Figure \ref{fig:apodproflab} shows both the measured apodizer profile and the designed apodization. The residual difference between the two curves is due to the finite size of the pinhole (the beam before the apodizer is slightly apodized, so the measured profile is slightly too bright in the center) and the finite angular resolution of the pupil re-imaging (the sharp edge of the apodizer is blurred).

\subsection{Focal plane and Lyot masks}
\label{sec:masks}
\begin{figure}[htb]
\includegraphics[scale=0.7]{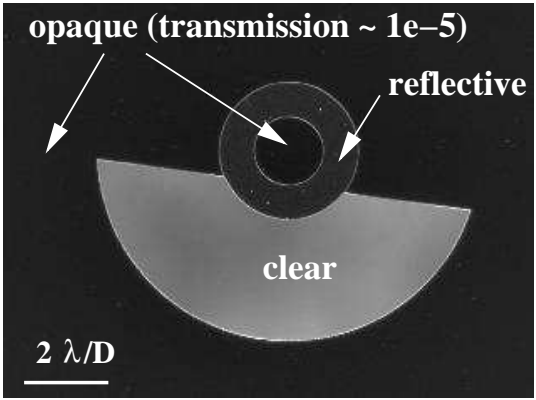} 
\caption{\label{fig:fpmask} Microscope image of the focal plane mask}
\end{figure}

The focal plane mask, shown in Figure \ref{fig:fpmask}, is used in transmission. The central part of the mask blocks the bright central PSF core. The radius of this non-transmissive central zone defines the inner working angle of the coronagraph, which is 1.65 $\lambda/D$ in our experiment. A clear zone transmits the science field to the science camera. The shape of this clear zone of the focal plane mask is chosen to exclude regions of the focal plane where the wavefront control system cannot remove diffracted light. Since our experiment uses a single deformable mirror, diffracted light can only be controlled over half of the field of view. The clear opening in the mask is therefore D-shaped with an outer radius imposed by the DM actuator sampling. A slightly larger rectangular zone could also have been adopted, but would have imposed the rotation angle of the focal plane mask. The reflective ring, extending from 0.8 $\lambda/D$ to 1.65$\lambda/D$, sends some of the starlight to the low order wavefront sensor (LOWFS) camera. We note that the opaque zones of the mask are not fully opaque due to manufacturing considerations: their transmition and reflection are respectively $\approx 10^{-5}$ and $\approx$ 10\%. 

A Lyot mask is located in the pupil plane between the focal plane mask and the science focal plane. This mask is designed to block all light outside the geometrical pupil and transmit all light within the pupil. It therefore has no effect on the nominal system throughput, and its role is to ensure that the scattered light reaching the focal plane camera does not contain light outside the pupil. Although correcting for such light is theoretically possible if it is coherent and within the control radius imposed by actuator sampling on the DM, it requires an accurate model of the coronagraph which can predict how light outside the pupil is affected by DM actuator positions. The Lyot mask was made by drilling a small hole in an aluminum plate, and its diameter is slightly smaller than the pupil size to account for alignment tolerances.

Both the focal and Lyot masks are on motorized stages and can be removed from the beam.

\section{Wavefront Control}
\label{sec:wfc}

As described in section \ref{wfc:hardarch}, wavefront control in our experiment is performed after the PIAA optics, which allows full decoupling between the pupil remapping introduced by the PIAA optics and the wavefront control algorithms. The wavefront calibration and wavefront control routines used in our experiment are therefore not specific to PIAA - they simply take as an input the apodized beam from the PIAA optics. We present in this section these routines as they are an essential part of the experiment, and the calibration techniques developed for this experiment are used to quantify the coronagraph's performance beyond the raw contrast achieved. Section \ref{ssec:wfc_initcalib} describes how the wavefront is first flattened and the DM response is calibrated in a non-coronagraphic mode (no focal plane mask). Section \ref{ssec:lowfs} briefly describes the low order wavefront control loop, which is detailed in a separate paper. In \S\ref{ssec:howfs}, the main wavefront control loop is described along with the calibrations used to measure the coronagraph performance beyond the raw contrast.

\begin{figure}[htb]
\includegraphics[scale=0.37]{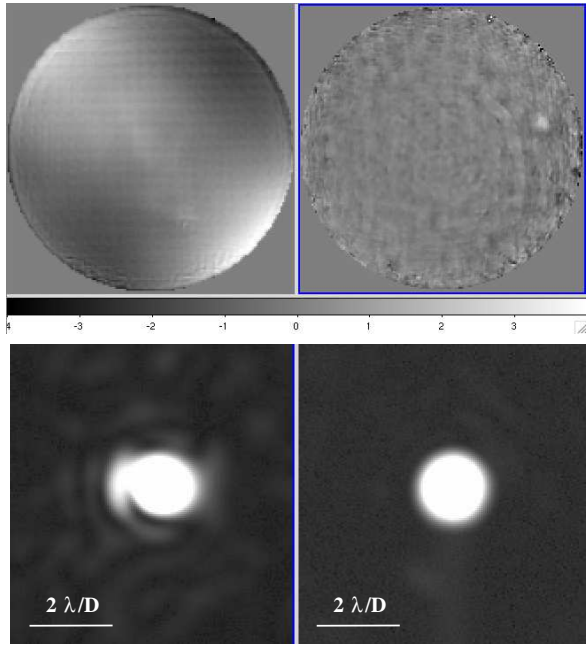} 
\caption{\label{fig:dmonoff} Pupil phase (top) and focal plane image (bottom) when the deformable mirror is powered off (left) and set to its nominal position after calibration (right). A malfunctionning actuator is visible on the right side of the beam.}
\end{figure}

\begin{figure}[htb]
\includegraphics[scale=0.45]{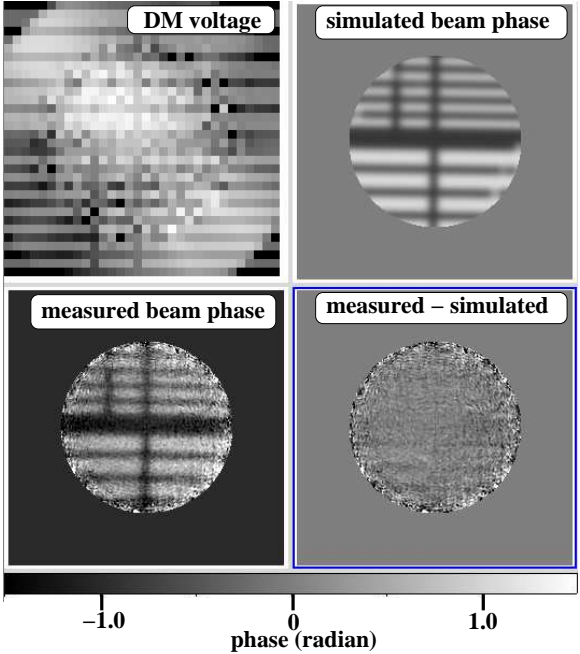} 
\caption{\label{fig:dmcalib} Result of the automatic DM calibration. The DM is driven to produce a recognizable pattern on top of the flat wavefront (DM voltages, top left). The DM model is used to convert the voltages into a simulated beam phase (top right). The difference between the measured beam phase (bottom left) and the simulated beam phase is shown in the bottom right panel. All images except the DM voltage map are shown with the same linear phase scale.}
\end{figure}

\subsection{Initial Calibration Loop (without focal plane mask)}
\label{ssec:wfc_initcalib}
Initial calibration is performed using conventional phase diversity with no focal plane mask: images are acquired with the science camera in six positions regularly spaced from the focal plane to the pupil plane. An iterative Gerchberg-Saxon algorithm is used to reconstruct the pupil plane complex amplitude. As shown in Figure \ref{fig:dmonoff} (top left), the beam quality is intially quite poor, with a large amount of astigmatism. The corresponding focal plane image is shown in Figure \ref{fig:dmonoff}, bottom left.

The phase diversity routine described above is reapeated $N$ times ($N \approx 10$), with a different set of DM voltages applied for each phase diversity measurement sequence. The $N$ phase maps obtained and the $N$ DM voltage maps used to obtain them are then used to constrain a model of the DM response which consists of seven parameters: Geometrical correspondance between the DM and the pupil image (4 parameters: x and y shift, scale, and rotation) and physical constants describing the DM behavior (3 parameters: width of the actuator influence function, DM dispacement for 100V applied and power index $\alpha$ in the displacement to voltage relationship with displacement $\propto$ $V^\alpha$).
The result of the DM calibration can then be used to flatten the wavefront measured and produce a sharp focal plane image (Figure \ref{fig:dmonoff}, right).

This initial calibration is a necessary preliminary step for the high contrast wavefront control, which needs (1) a knowledge of the starting point (typically less than 1 radian error on the wavefront) and (2) a good understanding of how DM commands affect the pupil plane phase. The quality of the DM calibration is shown in Figure \ref{fig:dmcalib} for a large DM offset (60 nm RMS in this example). The difference between the measured and simulated beam phase binned to the actuator size is 6 nm RMS in the central 75\% radius of the pupil (phase measurement in the outer part of the pupil is noisier due to lack of flux), corresponding to a 10\% relative accuracy. The DM model relative accuracy is better for smaller displacements.

\subsection{Low order wavefront errors}
\label{ssec:lowfs}
Low order wavefront errors are measured by reflecting a portion of the bright starlight masked by the coronagraph focal plane into a dedicted camera. A detailed description of this low order wavefront sensor (LOWFS) can be found in \cite{guyo09}. The LOWFS signal is used to simultaneously drive the deformable tip-tilt and the source position ahead of the PIAA optics. A key feature of the LOWFS is the ability to separate pointing errors (pre-PIAA tip-tilt) from post-PIAA tip-tilt, which is essential to maintain high contrast: even a small pre-PIAA tip-tilt creates diffraction rings outside the IWA of the coronagraph, and pre-PIAA tip-tilt errors cannot be compensated for by post-PIAA tip-tilt. 

When the low-order loop is closed, the measured residual pointing error is $10^{-3} \lambda/D$, and is therefore small enough to be negligible in the scattered light error budget shown in \S\ref{sec:coroperf}. A more detailed description of the design, calibration, control algorithm and performance of the LOWFS in our experiment is given in \cite{guyo09}.

\begin{figure*}[htb]
\includegraphics[scale=0.63]{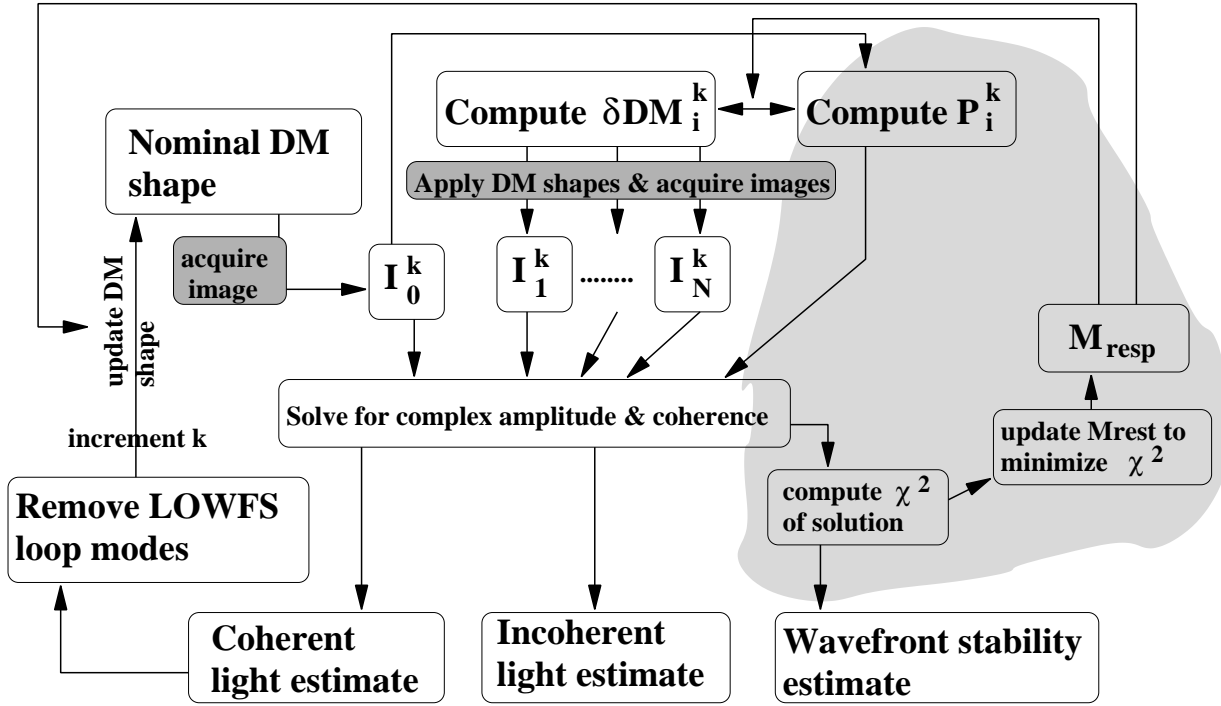} 
\caption{\label{fig:fpaowfcalgo} High order wavefront control loop, showing both the main loop and the system response matrix optimization loop (light shaded area). The two dark shaded boxes indicate image acquisition, which in the simulation mode, can be replaced with a simulated image acquisition using a model of the experiment and DM response.}
\end{figure*}

\subsection{High order wavefront control loop}
\label{ssec:howfs}
Coherent scattered light in the clear opening of the focal plane mask is measured by phase diversity introduced on the DM. A series of focal plane images, each acquired with a slightly different DM shape, is used to reconstruct the complex amplitude and coherence of the scattered light. The high order wavefront control loop uses a linearized representation of the system in focal plane complex amplitude, as described in the electric field conjugation (EFC) approach proposed by \cite{give07}. The wavefront control loop is shown in Figure \ref{fig:fpaowfcalgo}, and is built around the EFC approach.

Prior to starting the loop, a model of the coronagraph is used to compute how each actuator motion affects the complex amplitude in the focal plane. Since this relationship is linear for small displacements, this model is stored as a complex amplitude system response matrix $M_{resp}$ (shown on the right hand side of Figure \ref{fig:fpaowfcalgo}) of size $n$ by $m$, where $n$ is the number of DM actuators (ignoring actuators outside the pupil) and $m$ is the number of pixels in the high contrast region of the focal plane. $M_{resp}$ is the linear operator which establishes the relationship between deformable mirror actuator displacements $\delta DM(u,v)$ (sampled in $n$ points) and the corresponding complex amplitude change $\delta f(x,y)$ (sampled in $m$  points) in the focal plane:
\begin{equation}
\label{equ:Mresp}
\delta f(x,y) = M_{resp} \delta DM(u,v).
\end{equation}
$M_{resp}$ is therefore computed by moving each actuator of the DM in the coronagraph model and storing the corresponding change in focal plane complex amplitude in a column of $M_{resp}$.

\subsubsection{Loop initialization}
For each iteration $k$ of the loop, the first step in the wavefront sensing process is to acquire an image $I_0^k$ with the DM shape set at the best known position for high contrast imaging (upper left corner of Figure \ref{fig:fpaowfcalgo}). This first image is then  used to choose the shapes to apply on the DM to optimize measurement accuracy and sensitivity.

\subsubsection{Considerations for the choice of wavefront sensing DM shapes}
Residual light in the coronagraphic focal plane is measured by adding N known wavefront errors on the nominal deformable mirror shape. For each wavefront error added, a coronagraphic image is taken. Intensity variations between the N images encode the complex amplitude of the residual light that needs to be measured and removed. In this section, we discuss how to choose the N wavefront errors which are sent to the DM for sensing.

These N DM displacements are denoted $\delta DM_i^k(u,v)$, with i = 1 .. N (the index k denotes the wavefront control loop iteration). The complex amplitude added to the focal plane by each of the DM displacements is given by equation \ref{equ:Mresp}:
\begin{equation}
\label{equ:P_Mresp_DM}
P_i^k(x,y) = M_{resp} \delta DM_i^k(u,v).
\end{equation}
These complex amplitude functions are referred to as wavefront sensing probes. The amplitude of the probes must be carefully chosen: if the probes are too strong, the measurement is too sensitive to errors in the DM calibration; if they are too weak, the measurement is contaminated by photon noise, readout noise and small variations in the incoherent scattered light. As a guideline, it is therefore best to choose these DM offsets so that the additional light (the complex amplitude focal plane probes) is approximately as bright as the light which needs to be measured. Finally, randomly modulating the probes can mitigate the effect of calibration errors. 

\subsubsection{Wavefront sensing probes}

The first probe $P_1^k$ is chosen to satisfy, for each pixel $(x,y)$:
\begin{equation}
\label{equ:probe1}
|P_1^k(x,y)|^2 = \alpha_0 + \alpha_1 I_0^k(x,y)
\end{equation}
where $I_0^k(x,y)$ is the image acquired with DM shape $DM_0^k$. If $\alpha_1 = 0$, this constraint will force the DM shape to add a uniform coherent background of contrast $\alpha_0$ in the focal plane, while if $\alpha_0 = 0$ and $\alpha_1 = 1$ it will drive the DM shape to add a speckle map with the same intensity as in the $I_0^k$ image. The phase of this probe is not constrained, and is chosen by an iterative scheme to best satisfy equation \ref{equ:probe1}, with $\delta DM_1^k$ as the free parameter. We note that an exact solution to equation \ref{equ:probe1} may not exist within the DM space, but the following algorithm yields a good approximate solution:
\begin{enumerate}
\item{Initialization: $P_1^k$ is computed from equation \ref{equ:probe1} with a zero phase.}
\item{Singular value decomposition (SVD) of $M_{resp}$ is used to estimate $\delta DM_1^k$ from $P_1^k$. Eigenvalues close to zero are rejected in the SVD to improve stability.}
\item{$\delta DM_1^k$ is clipped to avoid large DM displacements.}
\item{Equation \ref{equ:P_Mresp_DM} is used to recompute $P_1^k$ from the clipped $\delta DM_1^k$.}
\item{The amplitude of $P_1^k$ is updated to satisfy equation \ref{equ:probe1}, but its phase is left unchanged from Step 4.}
\item{Return to Step 2 until the iterative algorithm is stopped.}
\end{enumerate}
This iterative algorithm produces simultaneously the probe $P_1^k$ and the corresponding DM displacements.

The second probe is chosen so that, at each point $(x,y)$ in the focal plane, its amplitude is identical to the first probe, but its phase is offset by $\pi/2$:
\begin{equation}
\label{equ:probe2}
P_2^k(x,y) \approx i \: P_1^k(x,y) 
\end{equation}
This $\pi/2$ phase offset maximizes the WFS sensitivity if the dominant sources of noises are photon noise and readout noise \citep{guyo05}. We note that if all DM actuators are functioning, there is a perfect solution to this equation, which can be obtained by shifting each spatial frequency of the $DM_1^k$ map by $\pi/2$. Images acquired with these first two probes, together with the image $I_0^k$, would be sufficient to solve for wavefront errors if light in the focal plane is fully coherent, but at least one more probe is needed to unambiguously measure light coherence, and more probes can also provide the redundancy required for implementation of the diagnostic tools described in \S\ref{sssec:wfcsolve} and \S\ref{sssec:respmopt}.

Two additional probes have been chosen to be $P_3^k = -P_1^k$ and $P_4^k = -P_2^k$, with exact solutions $\delta DM_3^k = -\delta DM_1^k$ and $\delta DM_4^k = -\delta DM_2^k$ respectively.

In our laboratory experiment, we chose to also add 5 more probes with random uncorrelated DM shapes of similar amplitude than the DM displacements obtained for probes 1 to 4 above. The mean dispacement amplitude per actuator is measured on probes 1 to 4, and random displacement maps are produced with the same amplitude. These additional probes are not required for wavefront reconstruction (probes 1 to 4 provide sufficient information), but, as described in the next sections, are added to allow for calibration of $M_{resp}$, measurement of incoherent light in the system, and measurement of coherent light variation during the sequence of $N$ exposures.

\subsubsection{Solving for complex amplitude and coherence}
\label{sssec:wfcsolve}
For each pixel $(x,y)$ of the focal plane detector, the complex amplitude $A(x,y)$ of the coherent light leak and the intensity $I(x,y)$ of the incoherent light leak is estimated by solving the following set of equations:
\begin{equation}
\label{equ:wfcsolve}
I_i^k(x,y) = |A(x,y)+P_i^k(x,y)|^2 + I(x,y)
\end{equation}
for $i=0 ... N$, with $P_0^k(x,y) = 0$ (image acquired with nominal DM shape). 

This set of equations has three unknowns and is therefore overconstrained for $N>2$. With $N=9$ adopted in our experiment, $A(x,y)$ and $I(x,y)$ are chosen to minimize :
\begin{equation}
\label{equ:chi2}
\chi^2(x,y) = \sum_{i=1..N} \left(I_i^k(x,y) - \left(|A(x,y)+P_i^k(x,y)|^2 + I(x,y)\right)\right)
\end{equation}
In noiseless ideal simulations, the residual $\chi^2$ should be null. The residual $\chi^2$ in the laboratory experiment includes errors due to:
\begin{itemize}
\item{Photon and readout noise in the $I_i^k(x,y)$ measurements}
\item{Variations in the light leaks during the measurement sequence. When solving for this set of equation, we assume $A(x,y)$ and $I(x,y)$ are static, but if they vary, $\chi^2$ will increase.}
\item{Systematic model errors (errors in $M_{resp}$), which lead to errors in the estimation of the values of $P_i^k(x,y)$. If $M_{resp}$ is wrong, then the DM command sent will not produce the expected $P_i^k(x,y)$.}
\end{itemize}

For convenience, we have scaled $\chi^2$ in coronagraphic contrast unit. This scaling is performed by measuring the uncorrelated noise that would need to be added to $A(x,y)$ between frames to reproduce the observed value of $\chi^2$. In this unit, the observed $\chi^2$ is approximately $4.5\:10^{-8}$ (see \S\ref{sec:labresults}).

The first contribution (photon + readout noise) has been computed to be a small part of the $\chi^2$ observed. We observed that increasing $\alpha_1$ (see equation \ref{equ:probe1}) above 1 has little effect on the residual $\chi^2$, also independantly suggesting that $\chi^2$ is not due to detector readout noise.

\subsubsection{System response matrix optimization}
\label{sssec:respmopt}
One key output of the wavefront control loop is the estimation of coherent light leaks (which should be used to compute the DM correction to apply for the next iteration), incoherent light leaks (which the DM can do nothing about) and the measurement of the wavefront stability during the measurement sequence. All these quantities depend upon a reliable estimation of $M_{resp}$. It is for example possible, if $M_{resp}$ is wrong, to obtain a low value of the residual coherent light and think the system has converged to a good contrast value, while in fact a significant amount of coherent light remains. This last issue is mitigated, but not entirely addressed, by continuously varying the probes (as this error is a function of the probes chosen, and will average to zero in the linear approximation used in this work if the probes are randomly chosen). An error in $M_{resp}$ would first appear as a large $\chi^2$ value for the solution of equations \ref{equ:wfcsolve}.

To address this, we have added a $M_{resp}$ optimization loop within our control loop. For each iteration $k$, the derivative of:
\begin{equation}
\chi^2 = \sum_{x,y}\chi^2(x,y)
\end{equation}
with each element of $M_{resp}$ is computed (this is a total of $2 \times n \times m$ derivatives, as the derivative is computed for the real and imaginary parts of each element of the $M_{resp}$ matrix). This derivative is computed from equation \ref{equ:chi2} by replacing $P_i^k$ by $M_{resp} \delta DM_i^k$ (equation \ref{equ:P_Mresp_DM}).

At each iteration of the wavefront control loop, $M_{resp}$ is then slightly modified in order to reduce $\chi^2$: for the 10\% of $M_{resp}$ elements showing the largest derivative against $\chi^2$, the value of the matrix element is moved in the direction indicated by the derivative by 0.1\% of the RMS value of elements in the corresponding column of $M_{resp}$. This algorithm was first tested on simulated data with an initial $M_{resp}$ estimate which was different from the actual $M_{resp}$ used in the simulation for computing the images. This test showed that $M_{resp}$ did converge toward the true $M_{resp}$, and that the $\chi^2$ value decreases as a result. Convergence is very slow, due to the large number of coefficients in the $M_{resp}$ matrix, requiring several hundred iterations before a significant improvement in $\chi^2$ is observed.

\subsubsection{Correction applied to the DM}
The linear electric field conjugation (EFC) algorithm \citep{give07} is used to cancel coherent scattered light. This algorithm uses the linearized coronagraph model which is also used for the measurement step described above. The system response matrix is inverted to build a control matrix which is multiplied to the coherent light estimate to yield the DM shape offset to be applied. Regularization schemes proposed by \cite{give07} were used to improve the loop stability and convergence speed.

\section{Laboratory results}
\label{sec:labresults}


\subsection{Alignment and Apodization measurement in the pupil plane}
The pupil plane apodization map is measured by placing the science camera in the pupil plane. In our experiment, the pupil is conjugated simultaneously to the PIAA M2 mirror, the apodizer, and the DM. Alignment is necessary to ensure that these three planes are conjugated and that their relative scales are correct. The camera positions for which conjugation to these planes is achieved are measured and the corresponding pupil scales are derived from the images. These six numbers are then fed to an optimization routine which computes the offsets to be applied to all movable optical elements after PIAA M2 to meet the conjugation and scale requirements. A few iterations of this sequence were sufficient to converge.

\begin{figure}[htb]
\includegraphics[scale=0.31]{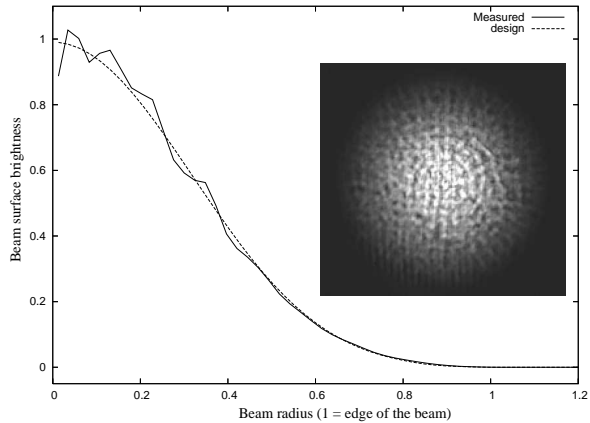} 
\caption{\label{fig:pupilIplot} Measured apodization profile, compared the to the beam profile the experiment was designed to deliver. A 2-D image of the beam is also shown.}
\end{figure}

In the fine alignment step, the apodizer alone is moved. The pupil image is compared to a simulated pupil image where the relative scale and lateral offset between the PIAA apodization and the apodizer transmission map are free parameters. The values of these three parameters which give the smallest residual difference is then used to guide fine alignment of the apodizer. Fine tuning of the scale between the apodizer and the PIAA apodization is possible because the beam at the apodizer is non collimated: apodizer motion along the optical axis changes this scale. Figure \ref{fig:pupilIplot} shows both the pupil apodization profile measured after alignment and the theoretical profile. 

\subsection{Imaging with a non-corrected PIAA system}

\begin{figure}[htb]
\includegraphics[scale=0.47]{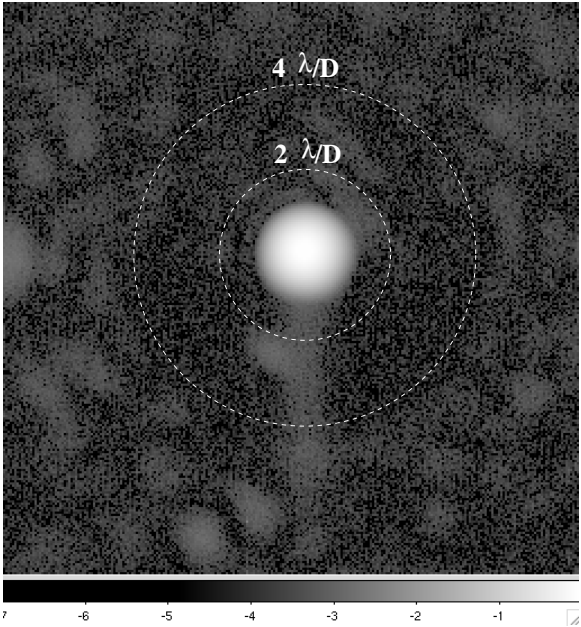} 
\caption{\label{fig:onaxispsf} On-axis PSF without focal plane mask (log scale indicated at the bottom of the figure). The vertical bleeding feature extending downward of the PSF core is a camera artifact, and is removed when the focal plane mask is used. A faint ghost due to the entrance window of the camera is visible just beyond 2 $\lambda/D$. }
\end{figure}

Figure \ref{fig:onaxispsf} shows the system on-axis PSF in imaging mode (no coronagraph focal plane mask). The on-axis PSF is similar to an Airy function without the Airy rings beyond 1.22 $\lambda/D$.

While the on-axis image is sharp and exibits high contrast, our laboratory system did not include the inverse PIAA optics necessary to correct for the strong off-axis aberrations introduced by the forward PIAA optics. These inverse optics do not need to be of coronagraphic quality, and can be a small set of lenses. A laboratory demonstration of wide field correction with inverse PIAA optics is described in a separate paper \citep{lozi09}.

\begin{figure}[htb]
\includegraphics[scale=0.37]{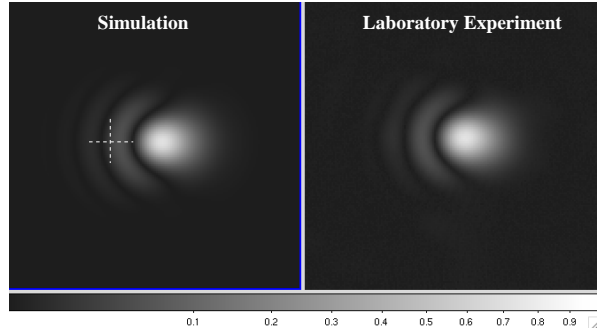} 
\caption{\label{fig:offaxiscomp} Off-axis PSF without focal plane mask, for a 1.7 $\lambda/D$ off-axis distance. A cross indicates the optical axis in the simulated image (left). Both images are shown to the same brightness scale (bottom).}
\end{figure}

Without inverse PIAA optics, the image of an off-axis source rapidly changes shape as the source moves away from the optical axis. Figure \ref{fig:offaxiscomp} shows the measured off-axis PSFs and the result of a simulations using a remapping of the beam phase \citep{guyo03}. With such a strong field aberration, measuring the focal plane plate scale is challenging and its value is a function of the metric used. In this paper, we choose to adopt the non coronagraphic PSF photocenter to measure plate scale. In Figure \ref{fig:onaxispsf}, the 4 $\lambda/D$ ring therefore shows where the photocenter of the PSF would be if the source was 4 $\lambda/D$ from the optical axis of the entrance telescope. The bright PSF core at this separation would be slightly outside the ring, but the fainter assymetric diffraction arcs of the off-axis PSF would be inside the ring. The same photocenter metric is used to measure the angular sizes on the focal plane mask given in \S\ref{sec:masks}. The plate scale on the coronagraph camera was measured by 2-D fitting of measured and simulated off-axis PSFs at several angular separations.  

\subsection{Coronagraphic performance: measurements}
\label{sec:coroperf}


\begin{figure*}[p]
\begin{center}
\includegraphics[scale=0.7]{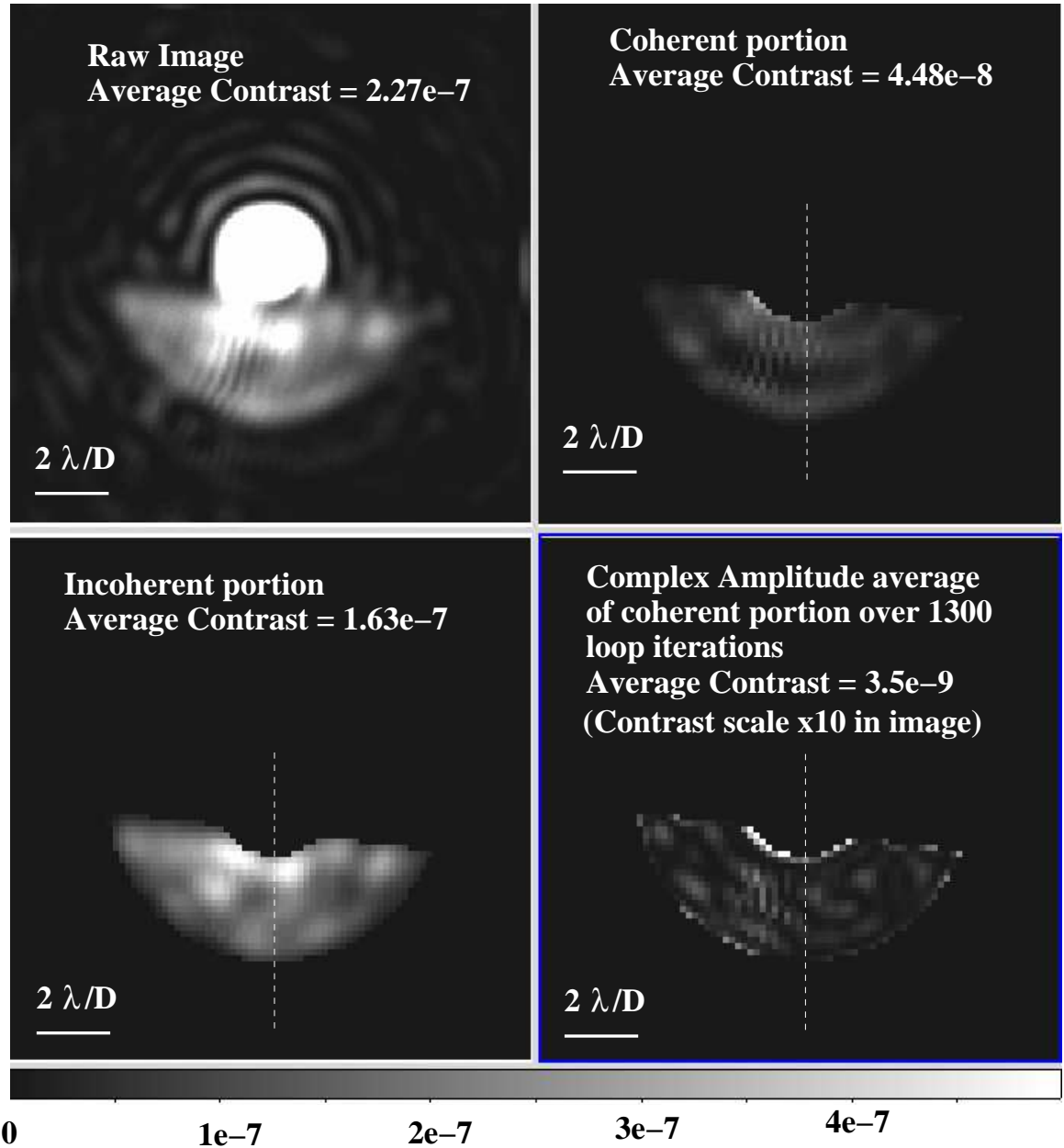} 
\end{center}
\caption{\label{fig:piaahighcontrastresult} Top left: Raw coronagraphic image. A decomposition of the scattered light into a coherent component (top right) and incoherent component (bottom left) shows that the raw contrast is dominated by incoherent light. The coherent bias over a long period of time is shown in the bottom right.}
\end{figure*}

Coronagraphic performance is measured with the focal plane and Lyot masks in the beam and both the LOWFS loop and high order wavefront control loop closed. Figure \ref{fig:piaahighcontrastresult} (upper left) shows a raw image from the science camera. Most of the light reaching the camera is due to partial transmittance (at the $10^{-5}$ level) of the core of the focal plane mask, which produces a central peak in the image. The clear $1.65 \lambda/D$ to $4.4 \lambda/D$ opening in the focal plane mask is visible below this central peak.

Residual light is decomposed in two components by the wavefront sensing sytem, using the approach described in \S\ref{ssec:howfs}:
\begin{itemize}
\item{An incoherent component composed of light which does not interfere with light extracted directly from the central PSF core. This component appears to be mostly stable in structure and varies from iteration to iteration between 1.5 $10^{-7}$ and 2 $10^{-7}$ in contrast.}
\item{A coherent component which is used to compute the DM shape for the next iteration. This component is at approximately $510^{-8}$ contrast and is decorrelated on timescales above the response time of the wavefront control loop. The estimation of this component varies greatly from iteration to iteration, and is sometimes below $10^{-8}$ contrast.}
\end{itemize}
We also measured the $\chi^2$ of the solution, and found it to correspond to a change in coherent light at the 4.5 $10^{-8}$ contrast level during the measurement sequence, which is 10s long. The $M_{resp}$ optimization loop described in \S\ref{sssec:respmopt} gave no noticable improvement in $\chi^2$, suggesting that the inital calibration led to a good estimate of $M_{resp}$ and that the observed $\chi^2$ is indeed dominated by fluctuations in coherent light at the 4.5 $10^{-8}$ contrast level. 

Visual inspection of the coherent and incoherent portions of the light strongly suggests that this decomposition was successful. First, the images obtained show very little high spatial frequency noise. Although the reconstruction is performed independantly for each pixel, both the coherent and incoherent components show speckles and features covering several pixels in these oversampled images. Second, the ghost to the lower left of the optical axis was properly analysed by our algorithm. This ghost, also visible in figure \ref{fig:onaxispsf}, is due to internal reflection in the window of the CCD camera. Because the two surfaces of the window are not perfectly parallel, interference fringes can be seen across this ghost. Although the fringes are coherent, they are too thighly spaced for the wavefront control system to remove. They are the only known feature in the image that (1) should be seen in the coherent image, (2) should not be seen in the incoherent image, and (3) cannot be removed by the wavefront control system. As shown in figure \ref{fig:piaahighcontrastresult}, our coherent/incoherent analysis correctly identifies these fringe as coherent, and only a minute fraction of the fringes is present in the incoherent portion of the image (likely due to small motion of the fringes during the measurement sequence).

\begin{deluxetable}{l c c c}
\tabletypesize{\small}
\tablecaption{\label{tab:contrasterrbudget} Contrast budget (Averaged over 1.6 to 4.4 $\lambda/D$)}
\tablehead{\colhead{Term} & \colhead{Likely origin} & \colhead{Value} & \colhead{Calibration}}
\startdata
Incoherent light, static & optical ghost, polarization & 1.63 $10^{-7}$ & Incoherent portion of WFS data\\
Coherent, variable in $t < 10s$ & turbulence, vibrations & 4 $10^{-8}$ & residual from WFS reconstruction\\
Coherent residual (in 10s) & & 5 $10^{-8}$ (typical) & Coherent portion of WFS data\\
                             & & 7 $10^{-9}$ (best) & \\
Coherent, static (in 4hr) & uncorrectable coherent light & $<$3.5 $10^{-9}$ & time-averaged coherent light\\
\enddata
\end{deluxetable}

\subsection{Analysis}

The analysis results are summarized in table \ref{tab:contrasterrbudget} and discussed in this section.

\subsubsection{Incoherent light component}
The raw contrast is dominated by a very stable incoherent component at 1.63 $10^{-7}$, which is likely due to ghosts and/or polarization missmatch. Given the high number of air-glass surfaces (twenty), including some which are not anti-reflection coated, a ghost at this level would not be too surprising. A possible source of polarization missmatch is the ring apodizer which has many 0.8 $\mu$m wide annular rings. The variations observed in the estimate of the incoherent light are due to the coherent light variation during the measurement, which affects the incoherent estimate. The data obtained is compatible with a fully static incoherent background, as would be expected from ghosts and polarization effects. We note that a fast varying coherent component could also be responsible for this light if it is varying sufficiently rapidly to appear incoherent within $\approx$ 1 second.

\subsubsection{Coherent light component}
During the measurement sequence, coherent light varies by $\approx$ 4.5 $10^{-8}$ in contrast due to turbulence or vibrations in the system, as shown by the $\chi^2$ analysis. The coherent light leak estimate over a 10 second period is $\approx$ 5 $10^{-8}$, which is at the level expected from the 4.5 $10^{-8}$ variations shown by the $\chi^2$ analysis. The large variation, from iteration to iteration, observed in the coherent light residual is due to the turbulence/vibrations in the system. We note that the lucky iterations where the 10 second averaged coherent light is estimated below $10^{-8}$ are artefacts of the time averaging during the measurement period: even during these lucky periods the coherent light did vary by $\approx$ 4.5$10^{-8}$ between the 10 frames of the measurement.

\subsubsection{Static wavefront aberrations}
The wavefront control loop successfully removes static coherent speckles. Over a 4hr period of time, we have measured the static coherent speckles to be at the 3.5 $10^{-9}$ contrast level. Except for a known ghost on the camera window, we could not detect any residual bias above this level in the residual time-averaged coherent light. 

The quality of the DM calibration (better than 10\% for small displacements) and the use of 10 probes chosen to avoid systematic bias ensures that within a single wavefront measurement, the coherent wavefront error is below $0.1 \times \sqrt{10} \approx 0.03$ times the DM probe amplitude. Since DM probes were chosen to be at the $10^{-7}$ contrast level, the statistical measurement error due to DM calibration errors is expected to be below the $10^{-7} \times 0.03^2 \approx 10^{-10}$ contrast level for each wavefront measurement. Decorrelation between wavefront measurements further reduces the effect of DM calibration errors on the static wavefront aberration measurement accuracy. DM calibration and other potential sources of errors are further reduced by the system response matrix optimization described in \S\ref{sssec:respmopt}. At the 3.5 $10^{-9}$ contrast level, the measurement of static wavefront aberration is therefore robust.

\section{Conclusion}
The results obtained in this experiment are especially encouraging for ground-based coronagraphy. The 2 $10^{-7}$ raw contrast we have achieved in the 1.65 to 4.4 $\lambda/D$ angular separation range already exceeds by two orders of magnitudes the raw contrast that can be hoped for in even a theoretically ideal Extreme-AO system \citep{guyo05}. We note that with a more careful optical design and anti-reflection coated optics, our experiment could probably have reached 5 $10^{-8}$ raw contrast (level of coherent light leak in the current experiment). More importantly, we have demonstrated that with the coronagraph + wavefront control architecture adopted in our experiment, static coherent speckles can be pushed down very low (3.5 $10^{-9}$) in long exposures. Our system successfully removed long term correlations in the coherent speckles, and their averaged level in long exposure was reduced with a $1/\sqrt{T}$ law. The combination of a high performance PIAA coronagraph and a focal-plane based wavefront control therefore appears extremely attractive for ground-based Extreme-AO. In that regard, our experiment has been a successful validation of the key technologies and control algorithms of the Subaru Coronagraphic Extreme-AO (SCExAO) system currently in assembly. The major differences between the SCExAO PIAA coronagraph and our laboratory prototype are (1) the need to design and operate a PIAA coronagraph on a centrally obscured pupil with thick spider vanes and (2) the need for corrective optics to recover a wide field of view. These two requirements have been validated in a separate laboratory experiment using the final SCExAO coronagraph optics \citep{lozi09}.

Our experiment was limited at the 2 $10^{-7}$ contrast by an optical ghost and at the 5 $10^{-8}$ contrast by turbulence or vibrations. The PIAA coronagraph could therefore not be tested to the contrast level required for direct imaging of Earth-like planets from space (approximately $10^{-10}$), although several key concepts were demonstrated, including simultaneous operation of a low-order wavefront sensor using starlight in the PSF core and high-order wavefront sensor using scattered light in the science focal plane. New calibration schemes which will be very useful for high contrast coronagraphy were also developed and validated, such as the system response matrix optimization loop, which can executed as a background task to fine-tune the system.

PIAA coronagraph technologies for high contrast space applications are now being actively developed and tested at NASA Ames Research Center and NASA Jet Propulsion Laboratory. A new set of PIAA mirrors was recently manufactured to higher surface accuracy than the ones used for this demonstration, and is being integrated within the High Contrast Imaging Testbed (HCIT) vacuum chamber at NASA Jet Propulsion Laboratory. We note that the HCIT chamber has already demonstrated stability to the $10^{-10}$ contrast with a Lyot-type coronagraph \citep{trau07}. The experiment described in this paper served as a precursor to this new step, which is aimed at reaching higher contrast (minimum goal of $10^{-9}$) in broadband light using a two-DM wavefront correction. In parallel to this effort, a highly flexible high stability air testbed at NASA Ames Research Center is coming online to further explore technology and architecture trades for PIAA systems.

In addition to pushing the contrast further, future laboratory demonstration of the PIAA coronagraph will explore chromaticity and wavefront control issues unique to a PIAA coronagraph architecture. The monochromatic experiment described in this paper did not address these important points, and should therefore be considered as only a first step toward validation of the PIAA coronagraph technique for high contrast space-based exoplanet imaging mission.

\acknowledgements
This research was conducted with funding from NASA JPL and the National Astronomical Observatory of Japan. Technical input and advice from the members of NASA Jet Propulsion Laboratory's High Contrast Imaging Testbed (HCIT) team, NASA Ames Research Center's coronaraph team, and Princeton University's coronagraph team have been of considerable help to conduct this work, both for design/simulations and laboratory implementation. In addition to providing laboratory space and infrastructure, Subaru Telescope made this research possible through major contributions from its technical staff (electronics, hardware, software).


\begin{thebibliography}{}
\bibitem[Belikov et al.(2006)]{beli06} Belikov, R., Kasdin, N.J., \& Vanderbei, R.J.  2006, \apj, 652, 833
\bibitem[Galicher et al.(2005)]{gali05} Galicher, R., Guyon, O., Otsubo, M., Suto, H., Ridgway, S.T. 2005, \pasp, 117, 411
\bibitem[Give'on et al.(2007)]{give07} Give'on, A. et al. 2007, \procspie, 6691
\bibitem[Guyon(2003)]{guyo03} Guyon, O.  2003, \aap, 404, 379
\bibitem[Guyon(2005)]{guyo05} Guyon, O, 2005, \aap, 629, 592
\bibitem[Guyon et al.(2005)]{guyo05} Guyon, O., Pluzhnik, E.A., Galicher, R., Martinache, F., Ridgway, S.T., \& Woodruff, R.A.  2005, \apj, 622, 744
\bibitem[Guyon et al.(2006)]{guyo06} Guyon, O.,  Pluzhnik, E. A., Kuchner, M. J., Collins, B., \& Ridgway, S. T.  2006, \apj, 167, 81
\bibitem[Guyon et al.(2006)]{guyo06} Guyon, O., Pluzhnik, E.A., Kuchner, M.J., Collins, B., Ridgway, S.T.  2006, \apj, 167, 81
\bibitem[Guyon et al.(2009)]{guyo09} Guyon, O., Matsuo, T., Angel, R.J.  2009, \pasp, in press
\bibitem[Lozi et al.(2009)]{lozi09} Lozi, J., Martinache, F., Guyon, O. 2009, submitted to \pasp
\bibitem[Martinache \& Guyon (2006)]{mart06a} Martinache, F., Guyon, O. 2006, EAS Publications Series, 22, 281
\bibitem[Martinache et al.(2006)]{mart06} Martinache, F., Guyon, O., Pluzhnik, E.A., Galicher, R., Ridgway, S.T. 2006, \apj, 639, 1129
\bibitem[Pluzhnik et al.(2006)]{pluz06} Pluzhnik, E.A., Guyon, O., Ridgway, S.T., Martinache, F., Woodruff, R.A., Blain, C., Galicher, R. 2006, 644, 1246
\bibitem[Traub \& Vanderbei(2003)]{trau03} Traub, W.A., Vanderbei, R.J.  2003, \apj, 559, 695
\bibitem[Trauger \& Traub(2007)]{trau07} Trauger, J.T., Traub, W.A.  2007, Nature, 446, 771
\bibitem[Vanderbei \& Traub(2005)]{vand05} Vanderbei, R.J., Traub, W.A. 2005, \apj, 626, 1079
\bibitem[Vanderbei(2006)]{vand06} Vanderbei, R.J.  2006, 636, 528
\end{thebibliography}
\end{document}